\documentclass[aps,twocolumn,showpacs,floats,prl]{revtex4}
\usepackage{graphicx}
\usepackage{dcolumn}
\usepackage{bm}
\usepackage{color}
\usepackage{amsmath}
\usepackage{subfigure}
\usepackage{enumitem}

\begin{document}

\author{L. A. Pe\~{n}a Ardila}
\affiliation{Max Planck Institute for the Physics of Complex Systems, N\"othnitzer Stra\ss e 38, 01187 Dresden, Germany}
\author{S. Giorgini}
\affiliation{Dipartimento di Fisica, Universit\`a di Trento and CNR-INO BEC Center, I-38123 Povo, Trento, Italy}

\title{The Bose polaron problem: effect of mass imbalance on binding energy} 

\begin{abstract} 
By means of Quantum Monte Carlo methods we calculate the binding energy of an impurity immersed in a Bose-Einstein condensate at $T=0$. The focus is on the attractive branch of the Bose polaron and 
on the role played by the mass imbalance between the impurity and the surrounding particles. For an impurity resonantly coupled to the bath, we investigate the dependence of the binding energy on the mass ratio and on the interaction strength within the medium. In particular, we determine the equation of state in the case of a static (infinite mass) impurity, where three-body correlations are irrelevant and the result is expected to be a universal function of the gas parameter. For the mass ratio corresponding to $^{40}$K impurities in a gas of $^{87}$Rb atoms we provide an explicit comparison with the experimental findings of a recent study carried out at JILA.
\end{abstract}

\maketitle

\section{I. Introduction and method}
\label{sec:intro}

In two recent experiments~\cite{JILA, Aarhus} the energy of an impurity immersed in a Bose-Einstein condensate (BEC) has been measured using RF spectroscopy and a Feshbach resonance to tune the impurity-bath interaction strength. The most striking result is the existence of a well defined quasiparticle peak which, on the attractive branch of the Bose polaron, extends up to the unitary point where the impurity is resonantly interacting with the medium. In the experiment carried out at JILA~\cite{JILA} a low density gas of fermionic $^{40}$K impurities is superimposed to a BEC of $^{87}$Rb atoms, whereas researchers in Aarhus have used two different hyperfine states of bosonic $^{39}$K atoms. Even though the qualitative results are similar in the two experiments, quantitative differences in the energy spectrum arise due to the different mass ratio, $m_B/m_I\simeq2$ in Ref.~\cite{JILA} and $m_B/m_I=1$ in Ref.~\cite{Aarhus} where $m_B$ is the mass of the particles in the BEC and $m_I$ is the mass of the impurity, as well as to the different gas parameter of the medium, respectively $na^3\simeq3\times10^{-5}$ and $na^3\simeq2\times10^{-8}$ in the JILA and Aarhus experiment being $n$ the density and $a$ the scattering length describing interactions within the BEC. Further differences, concerning in particular the width of the spectrum at resonance, probably arise from the coupling to the continuum of excited states which affects the measured response function~\cite{Aarhus}.

The fermionic analogue of this problem, {\it i.e.} an impurity immersed in a Fermi sea, has already been the object of many experimental and theoretical studies~\cite{Fermipolaron}. In particular, on the theory side, the Fermi polaron problem has been addressed using quantum Monte Carlo (QMC) methods which are best suited to deal with the regime of strong correlations between the impurity and the medium. The binding energy of the attractive polaron was calculated exactly by means of the diagrammatic Monte Carlo technique~\cite{Prokofev1, Prokofev2}, whereas projection Monte Carlo methods implementing the fixed-node approximation were used to calculate the equation of state of highly imbalanced Fermi mixtures both with equal~\cite{Pilati08} and unequal masses~\cite{Gezerlis09}. 

In a previous study~\cite{Ardila15} we characterized the properties of the Bose polaron using QMC methods in the case where the impurity and the particles in the BEC have the same mass. In particular, we determined the dependence of the binding energy of the impurity resonantly interacting with the bath by varying the gas parameter of the BEC. In the present work we extend the previous calculations by analyzing how this energy changes also as a function of the mass ratio $m_B/m_I$. We report on results ranging from the value $m_B/m_I=2$ of the light impurities used in the JILA experiment~\cite{JILA} to $m_B/m_I=0$ corresponding to the limit of a static impurity. This latter case is an interesting reference problem, which has an exact solution for an impurity in a Fermi sea~\cite{Recati07}, and for the Bose polaron is expected to have a universal dependence on the gas parameter of the bath. We also calculate the binding energy of the attractive branch for the parameters of the JILA experiment in order to allow for a direct comparison.  

We calculate the ground state of the following Hamiltonian
\begin{eqnarray}
H&=&-\frac{\hbar^2}{2m_B}\sum_{i=1}^N \nabla_i^2+\sum_{i<j}V_B(r_{ij}) 
\nonumber\\
&-&\frac{\hbar^2}{2m_I}\nabla_\alpha^2+\sum_{i=1}^NV_I(r_{i\alpha}) \;,
\label{eq:Hamiltonian}
\end{eqnarray}
describing a system of $N$ bosons of mass $m_B$ and one impurity of mass $m_I$. The particle coordinates are denoted by ${\bf r}_\alpha$ and ${\bf r}_i$ ($i=1,\dots,N$) for the impurity and the bosons. The pair potentials $V_B(r)$ and $V_I(r)$ denote, respectively, the boson-boson and the boson-impurity interaction and depend on the distance between the two particles. Similarly to Ref.~\cite{Ardila15}, we model $V_B$ using hard spheres with diameter $a$, which corresponds to the value of the inter-bosons $s$-wave scattering length. For the boson-impurity interaction $V_I$ we use a zero-range potential characterized by the scattering length $b$. The diffusion Monte Carlo (DMC) method used in the simulations has been already described in Ref.~\cite{Ardila15} and we refer the reader to this reference for details on the numerical technique. 

The mass-ratio dependence, arising from $m_I\neq m_B$, is accounted for in the DMC simulations by the properly corrected contribution of the impurity kinetic energy. Another slight modification compared to the method described in Ref.~\cite{Ardila15} concerns the trial wave function 
\begin{equation}
\psi_T({\bf r}_\alpha,{\bf r}_1,\dots,{\bf r}_N)=\prod_{i=1}^N f_I(r_{i\alpha})\prod_{i<j}f_B(r_{ij}) \;,
\label{trialwf}
\end{equation}
and arises from the zero-range nature of the impurity-boson potential. The boson-boson Jastrow correlation term $f_B$ coincides with the one used in Ref.~\cite{Ardila15}, whereas the impurity-boson function is given by
\begin{eqnarray}
f_I(r)=\begin{cases}
1-\frac{b}{r} & r<\bar{R} \\
B+C\left(e^{-\alpha r}-e^{-\alpha(L-r)}\right) & \bar{R}<r<L/2\;.
\end{cases}
\label{Jastrow}
\end{eqnarray} 
Here the parameters $B$ and $C$ are fixed by the continuity condition of the function $f_I$ and its first derivative at the matching point $\bar{R}$. This latter and the parameter $\alpha$ are optimized using a variational procedure on the total energy. It is important to stress that the Jastrow term (\ref{Jastrow}), which is singular in $r=0$, allows one to remove completely the interaction potential $V_I$ from the calculation of the energy and to substitute it with the contact boundary condition $f_I(r)=1-b/r$, fixed by the impurity-boson scattering length $b$. This procedure of replacing a short-range potential with the corresponding Bethe-Peierls boundary conditions has been already introduced in DMC simulations~\cite{Pessoa1, Pessoa2}, where also a particular sampling technique was implemented to handle the zero-range interaction. We find that the branching process in the DMC method~\cite{Mitas11}, by suppressing the weight of configurations with two or more bosons close to the impurity, reduces significantly the problem of large variance affecting variational calculations and allows one to carry out simulations without ``ad hoc'' sampling. We have checked that all the results reported in Ref.~\cite{Ardila15} for the square-well potential $V_I$ with the smallest range are faithfully reproduced by the choice (\ref{Jastrow}) of the trial wave function~\cite{Note1}.

\section{II. Results}
\label{sec:results}

The binding energy of the impurity is defined as the energy difference
\begin{equation}
\mu=E_0(N,1)-E_0(N) \:,
\label{bindener}
\end{equation}
between the ground states of the system with $N$ bosons and the impurity and of the clean system with only $N$ bosons. Both energies are calculated using the DMC method described in Ref.~\cite{Ardila15} for fixed values of $N$ in a box with periodic boundary conditions. The extrapolation to the thermodynamic limit is achieved by ensuring that $N$ is large enough to make finite-size effects negligible. To this aim calculations are repeated for different numbers $N$, typically $N=$ 32, 64 and 128. All energies are measured in units of the following convenient scale provided by the bath
\begin{equation}
\mu_0=\frac{\hbar^2(6\pi^2n)^{2/3}}{2m_B} \;,
\label{scale}
\end{equation}
which corresponds to the Fermi energy of a single-component Fermi gas with the same density and mass.

\begin{table*}
\centering
\caption{$\mu/\mu_{0}$ as a function of the mass ratio $m_{B}/m_{I}$ and the gas parameter $na^{3}$}
\begin{tabular}{|c|c|c|c|c|}
\hline 
$m_{B}/m_{I}$ & $na^{3}=5\times10^{-4}$ & $na^{3}=2.66\times10^{-5}$ & $na^{3}=1\times10^{-6}$ & $na^{3}=3\times10^{-7}$\tabularnewline
\hline 
$0.0$ & $-0.029(3)$ & $-0.35(5)$ & $-0.39(1)$ & $-0.44(3)$\tabularnewline
\hline 
$0.2$ & $-0.037(3)$ & $-0.33(3)$ & $-0.38(4)$ & $-0.34(7)$\tabularnewline
\hline 
$0.5$ & $-0.080(3)$ & $-0.47(4)$ & $-0.50(1)$ & $-0.47(1)$\tabularnewline
\hline 
$0.8$ & $-0.219(3)$ & $-0.63(6)$ & $-0.70(1)$ & $-0.73(3)$\tabularnewline
\hline 
$1.0$ & $-0.317(4)$ & $-0.81(5)$ & $-0.99(5)$ & $-1.06(1)$\tabularnewline
\hline 
$1.5$ & $-0.600(1)$ & $-1.21(2)$ & $-1.74(6)$ & $-2.1(1)$\tabularnewline
\hline 
$2.0$ & $-0.9    (2)$ & $-1.56(4)$ & $-2.4(5)$  &  $-4.4(4  )$\tabularnewline
\hline 
\end{tabular}
\label{tab1}
\end{table*}

\begin{figure}
\begin{center}
\includegraphics[width=8.5cm]{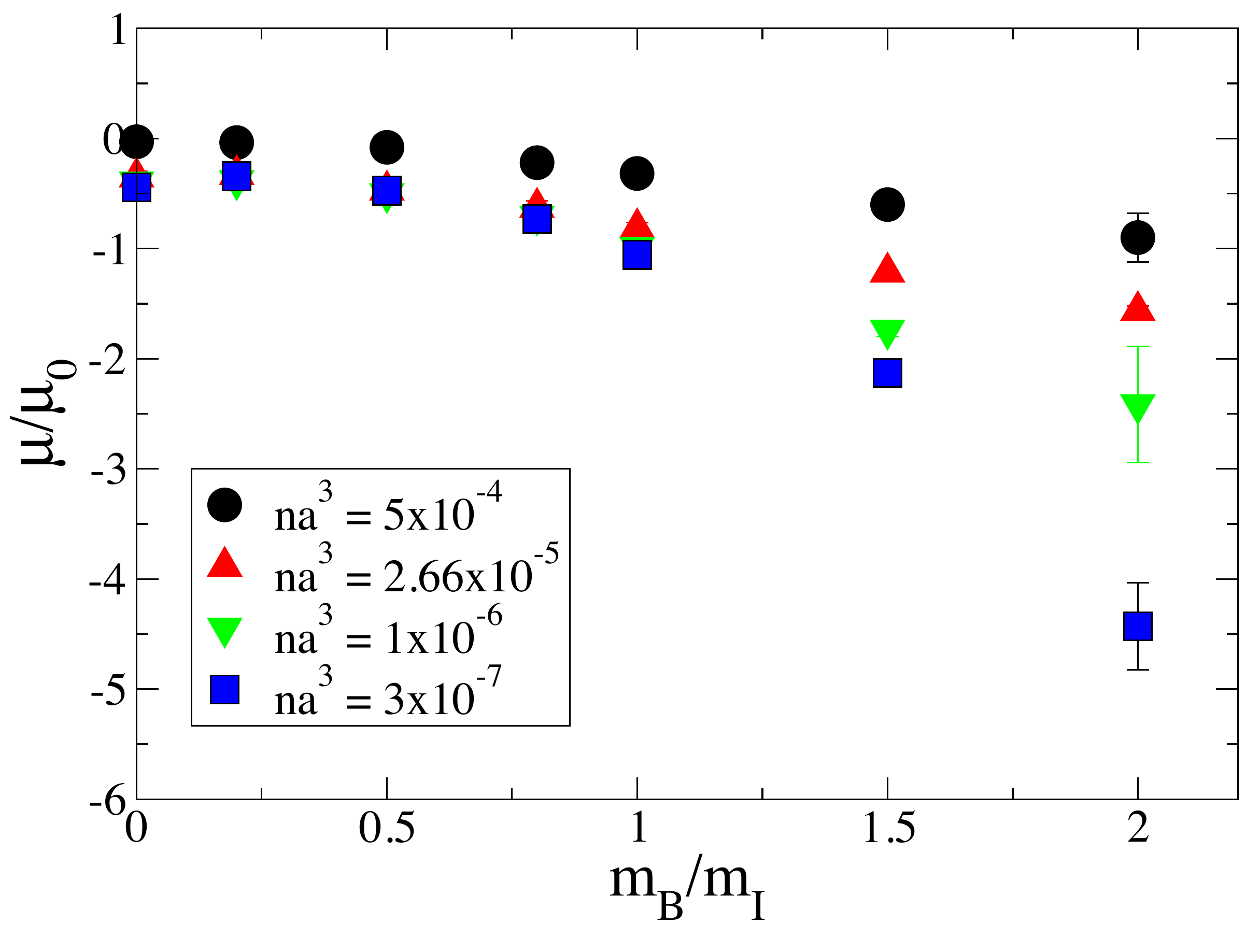}
\caption{(color online). Binding energy of the polaron for resonant coupling ($1/b=0$) as a function of the mass ratio $m_B/m_I$. Results corresponding to different values of the interaction strength in the bath are shown. In particular, the value $na^3=2.66\times10^{-5}$ refers to the conditions of the JILA experiment of Ref.~\cite{JILA}.}
\label{fig1}
\end{center}
\end{figure}

In Tab.~\ref{tab1} we report the results of the binding energy $\mu$ at the resonant value of the impurity-boson interaction ($1/b=0$) for different values of the mass ratio $m_B/m_I$ and of the gas parameter $na^3$. The same results are shown in Fig.~\ref{fig1}. The chosen values of the mass ratio range from $m_B/m_I=0$, corresponding to the case of a static impurity with infinite mass, to $m_B/m_I=2$. For lighter impurity masses ($m_B/m_I>2$) the calculation of $\mu$ becomes increasingly difficult because of instabilities towards the formation of cluster states. We notice that, at a given interaction strength within the bath, the binding energy increases in absolute value when the mass ratio increases. This effect is small for mass ratios up to $m_B/m_I=0.5$, while a larger drop of $\mu$ is seen when the impurity gets lighter than the particles in the medium.   

The case of a static impurity ($m_B/m_I=0$) resonantly coupled to the medium is particularly interesting. First of all, as visible from the results in Fig.~\ref{fig1}, it applies also to physically relevant situations where $m_I$ is significantly larger than $m_B$. Secondly, it is a reference problem which in the case of the equivalent Fermi polaron, {\it i.e.} a static impurity resonantly coupled to a single-component Fermi sea, features the exact solution $\mu/\mu_0=-1/2$~\cite{Recati07}. The bosonic counterpart is not exactly solvable, nevertheless, since Efimov physics is irrelevant due to the vanishing mass ratio~\cite{Petrov} and also tetramer and larger bound states should not play any role, one expects that $\mu/\mu_0$ is a universal function of the gas parameter in the BEC. 

\begin{figure}
\begin{center}
\includegraphics[width=8.5cm]{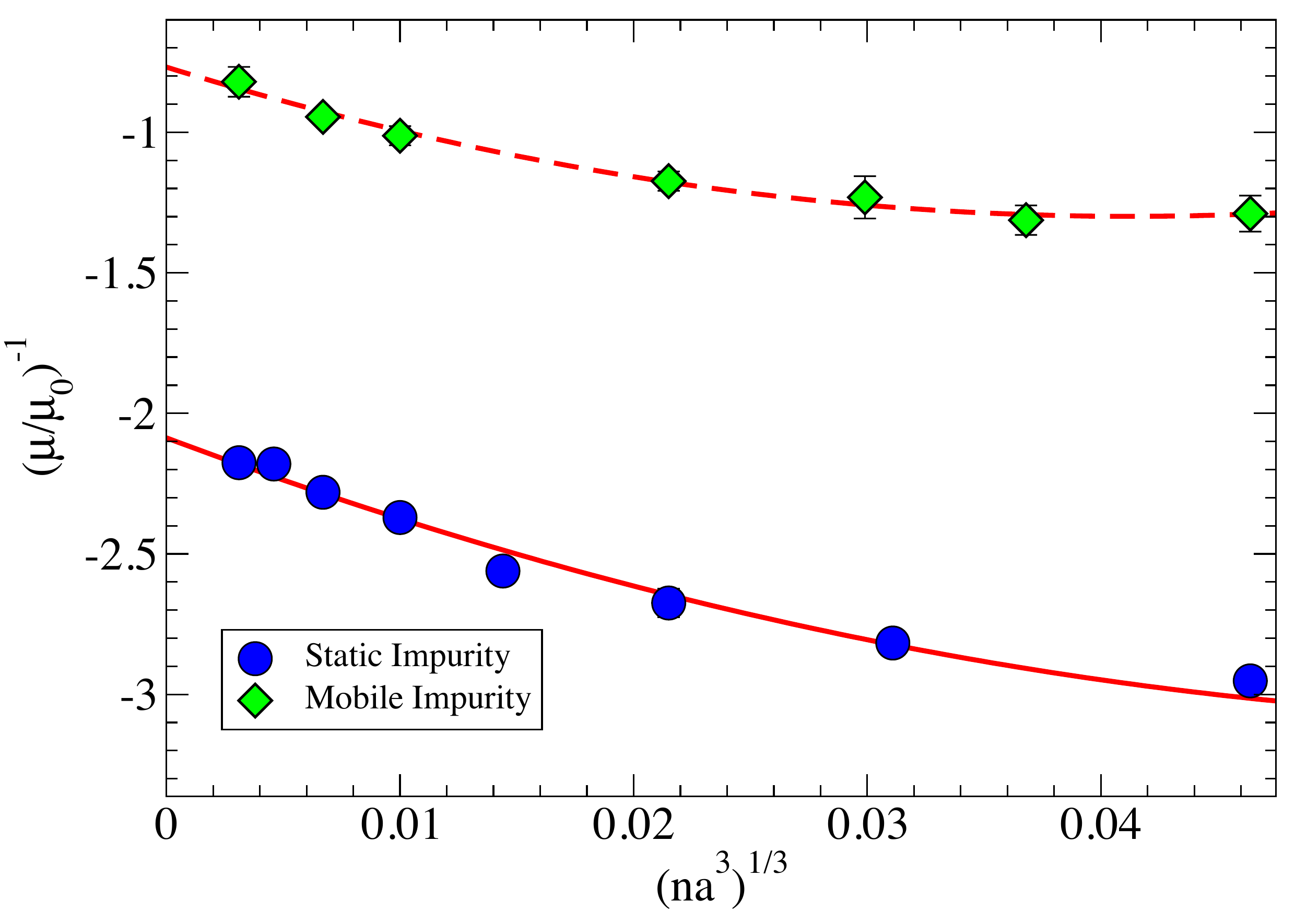}
\caption{(color online). Inverse binding energy of the polaron for resonant coupling ($1/b=0$) as a function of the gas parameter in the bath. Results are shown for a mobile impurity with $m_B/m_I=1$ (see Ref.~\cite{Ardila15}) and for a static impurity ($m_B/m_I=0$). The solid (static impurity) and dashed line (mobile impurity) are polynomial fits to the data including first and second order terms in $n^{1/3}a$.}
\label{fig2}
\end{center}
\end{figure}

The results for the inverse binding energy $(\mu/\mu_0)^{-1}$ are presented in Fig.~\ref{fig2} as a function of $n^{1/3}a$. One finds that $|\mu|$ increases with decreasing gas parameter, remaining however always smaller than the value 1/2 of the Fermi polaron mentioned above. In the same Fig.~\ref{fig2} we also report the corresponding results for the mass ratio $m_B/m_I=1$~\cite{Ardila15} which exhibit a similar trend as a function of $na^3$, but the value of $\mu$ is about a factor of two larger compared to the static impurity. Noticeably, the resonant Fermi polaron shows a much weaker dependence on the mass ratio as it reaches $\mu/\mu_0\simeq-0.58$ for equal masses~\cite{Lobo06}. 

One should also notice that, for both values of mass ratio in Fig.~\ref{fig2}, the dependence of the energy $\mu$ on the interaction strength $na^3$ in the bath is sizeable. Large variations are also predicted for the mass ratio $m_B/m_I=2$, as shown in Fig.~\ref{fig1}. We point out that such a strong dependence is completely underestimated in other theoretical approaches to the Bose polaron problem where interactions within the condensate are either completely neglected, as in the variational approach of Ref.~\cite{DasSarma14}, or produce only small effects, as in the T-matrix approximation of Ref.~\cite{Rath13}.  In Ref.~\cite{Bruun1}, instead, the effect of three-body correlations is investigated and a significant decrease of the polaron energy is found as a result of the avoided crossing between the quasiparticle and the deepest Efimov state.

We investigated the presence of cluster states, {\it i.e.} bound states of the impurity with two or more bosons, at resonant coupling both for the mass ratio $m_B/m_I=0$ and $m_B/m_I=1$. While in this latter case we find very shallow bound states of up to five bosons~\cite{Ardila15}, in the case of a static impurity no bound states are found within our model of hard-sphere boson-boson interactions. The role of such cluster states, and in particular of Efimov trimers, becomes important in the proximity of the resonance by adding new length scales to the problem and thereby suppressing the universal character of the polaron energy in terms of the gas parameter of the surrounding BEC~\cite{Bruun1}. We believe that our results for $m_B/m_I=1$ describe correctly the polaron ground state provided the energy of the deepest Efimov trimer is much smaller than $\mu_0$. 

We decided to plot the results for the binding energy in Fig.~\ref{fig2} in terms of $(\mu/\mu_0)^{-1}$ in order to discuss more properly the limit $na^3\to0$. For the static impurity at $1/b=0$, one can easily verify that an ideal Bose gas, featuring $a=0$, is unstable against collapse and yields a binding energy $\mu$ which scales with the $N^{1/3}$ law of the number of bosons~\cite{Note2}. Results in Fig.~\ref{fig2} show instead that $1/\mu$ does not scale to zero as $na^3\to0$, provided that one first takes the thermodynamic limit at fixed $na^3$ and only after allows the gas parameter to approach zero. The mobile impurity with mass ratio $m_B/m_I=1$ exhibits a similar behavior (see Fig.~\ref{fig2}). However, in this case, one expects that for small enough densities the energy $\mu$ approaches the binding energy of the deepest cluster state.   

\begin{figure}
\begin{center}
\includegraphics[width=8.5cm]{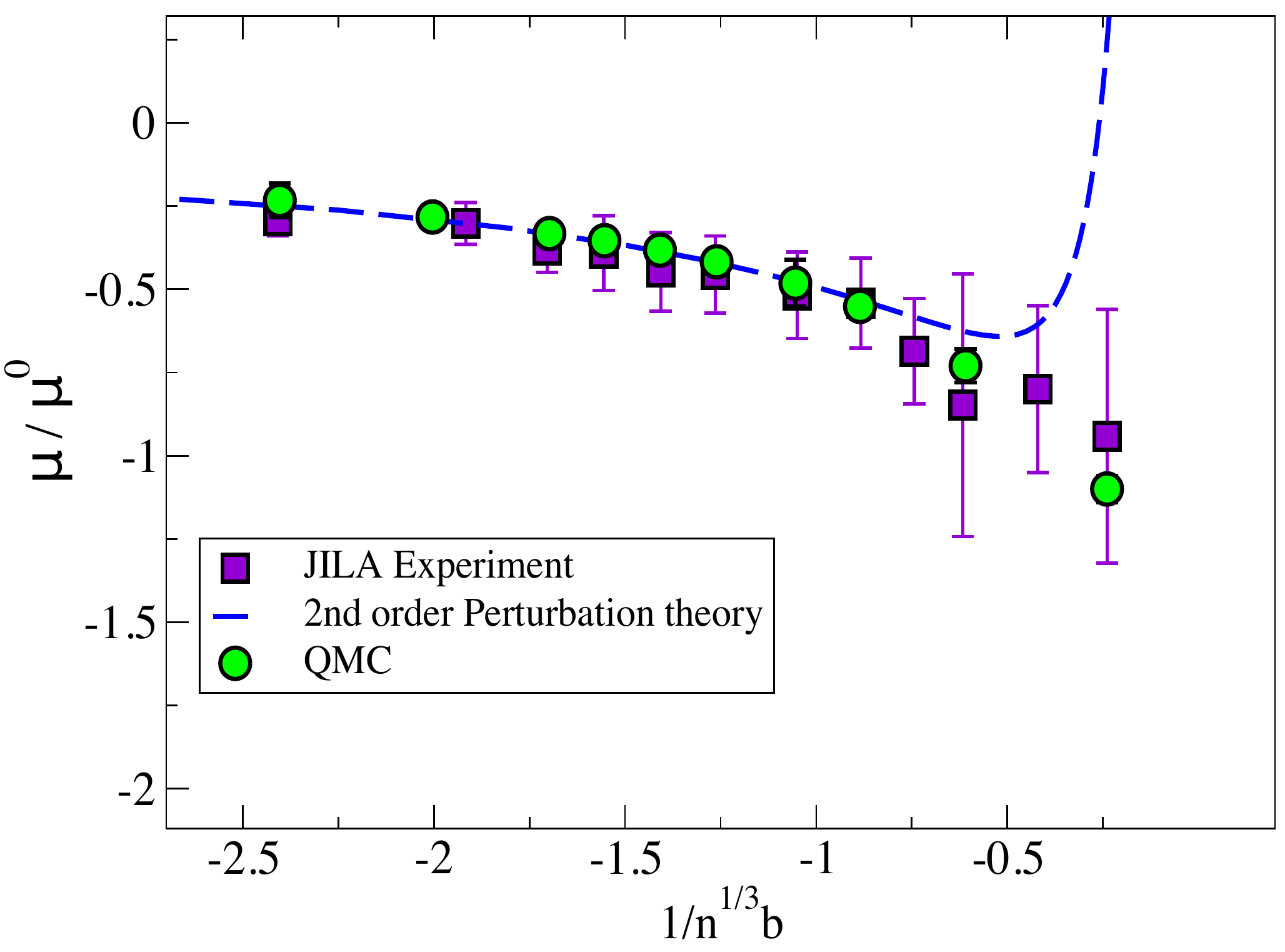}
\caption{(color online). Binding energy of the impurity along the attractive branch as a function of the impurity-bath coupling $1/(n^{1/3}b)$. The mass ratio is $m_B/m_I=2$ and is close to the case of $^{40}$K impurities in a BEC of $^{87}$Rb atoms as in the JILA experiment~\cite{JILA}. The gas parameter has also been chosen to reproduce the experimental value $na^3=2.66\times 10^{-5}$. The experimental points are reproduced from Fig.~(3) in Ref.~\cite{JILA}. The blue dashed line corresponds to the prediction of perturbation theory including first and second order terms. }
\label{fig3}
\end{center}
\end{figure}

Finally, in Fig.~\ref{fig3}, we compare the binding energy obtained from our DMC simulations with the results of the JILA experiment~\cite{JILA}. Both the value of the chosen mass ratio, $m_B/m_I=2$, and of the interaction strength, $na^3=2.66\times10^{-5}$, are close to the parameters of the experiment. The results cover the whole attractive branch of the Bose polaron with $b<0$ from the weakly interacting regime, where $|b|$ is on the order of the interparticle distance $n^{-1/3}$ and good agreement is found with the prediction of second-order perturbation theory~\cite{Note3}, to the resonant point achieved at $1/b=0$. The quantitative agreement between theory and experiment is good, although the large experimental uncertainty does not allow for a stringent comparison. In the case of the Aarhus experiment~\cite{Aarhus}, where $na^3\simeq2\times10^{-8}$ and $m_B/m_I=1$, our results would predict $\mu\simeq-1.2\mu_0$ at $1/b=0$, which is more than a factor two smaller than the measured value. A possible explanation of this discrepancy can come from the coupling to the continuum of excited states above the polaron ground state, which is crucial to give a proper account of the measured spectral response and produces a shift towards higher energies of the signal peak~\cite{Aarhus}. It is however unclear why such effects appear to be less important in the case of the JILA experiment, where the quasiparticle peak remains well defined up to the resonant point~\cite{JILA}.

In conclusion, we calculated the binding energy of the Bose polaron resonantly coupled to the surrounding bath as a function of the mass ratio and of the interaction strength within the medium, thereby extending the analysis of Ref.~\cite{Ardila15} where only the equal-mass case was considered. Furthermore, we determine the equation of state of a static impurity in terms of the gas parameter of the BEC. These results can be useful in understanding to what extent the properties of the Bose polaron are universal functions of the boson-impurity scattering length and of the strength of interactions in the medium. In the case of the recent JILA experiment~\cite{JILA}, our results for the corresponding value of the mass ratio are in good agreement with the measured polaron energy along the attractive branch.   

\section{Acknowledgements}

We gratefully acknowledge valuable discussions with G. M. Bruun and M. M. Parish. This work was supported by the QUIC grant of the Horizon2020 FET program and by Provincia Autonoma di Trento.

\end{document}